\documentclass[english,one column]{IEEEtran}
\usepackage[T1]{fontenc}
\usepackage[latin9]{inputenc}
\usepackage{geometry}
\usepackage{color}
\usepackage{array}
\usepackage{multirow}
\usepackage{amsthm}
\usepackage{amsmath}
\usepackage{amssymb}
\usepackage{graphicx}

\makeatletter

\providecommand{\tabularnewline}{\\}

\theoremstyle{plain}
\newtheorem{thm}{\protect\theoremname}
\theoremstyle{definition}
\newtheorem{defn}[thm]{\protect\definitionname}
\theoremstyle{plain}
\newtheorem{lem}[thm]{\protect\lemmaname}

\makeatother

\usepackage{babel}
\providecommand{\definitionname}{Definition}
\providecommand{\lemmaname}{Lemma}
\providecommand{\theoremname}{Theorem}

\begin{document}

\title{Secret-message capacity of a line network}

\author{$\vspace{-0.5cm}$
\begin{eqnarray*}
\textrm{Athanasios Papadopoulos\;\;\;\;}\;\; & \textrm{Laszlo Czap} & \textrm{\;\;\;\;\;\;\ Christina Fragouli}\\
\textrm{UCLA\;\;\;\;\;\;\;\;\;\;}\;\;\;\;\;\;\;\;\;\; & \textrm{EPFL} & \;\;\;\;\;\;\;\;\;\;\;\;\textrm{UCLA/EPFL}\\
\textrm{athanasios.papadopoulos@ucla.edu} & \textrm{laszlo.czap@epfl.edu} & \textrm{christina.fragouli@ucla.edu}
\end{eqnarray*}
}

\maketitle
$\vspace{-1.6cm}$

\section{Abstract}

We investigate the problem of information theoretically secure communication in a line network
with erasure channels and state feedback. We consider a spectrum of cases for the private randomness
that intermediate nodes can generate, ranging from having intermediate nodes generate unlimited
private randomness, to having intermediate nodes generate no private randomness, and all cases
in between. We characterize the secret message capacity when either only one of the channels
is eavesdropped or all of the channels are eavesdropped, and we develop polynomial time algorithms
that achieve these capacities. We also give an outer bound for the case where an arbitrary
number of channels is eavesdropped. Our work is the first to characterize the secrecy capacity
of a network of arbitrary size, with imperfect channels and feedback. As a side result, we
derive the secret key and secret message capacity of an one-hop network, when the source has
limited randomness.

\section{Introduction}

We consider a source that communicates with a destination over a line network with N edges,
where each intermediate node represents a relay, and each edge represents an erasure channel
with state feedback (all channels are assumed to be orthogonal). The source aims to send a
message securely to the destination, in the presence of a passive eavesdropper, Eve, who wiretaps
an (unknown) subset with (known) cardinality V of the N channels. Eve receives independently
erased versions of the transmissions, as well as the transmitted state feedback from all the
channels. We are interested in (strong) information theoretical secrecy.

We believe that the above setup is an interesting scenario for two reasons. First, line networks
capture single paths in arbitrary networks; indeed, today the vast majority of communication
occurs by connecting a source to a destination through a single path. Moreover, feedback is
an integral part of most communication protocols, making it possible to exploit it for secrecy.
Thus the setup we consider approaches current practices. Second, an understanding of the single
path is a necessary first step towards exact characterizations of more general networks.

The main contribution of this paper is to exactly characterize the capacity over an arbitrarily
long line network with erasures and feedback when V = 1 and V = N. A series of recent papers
in the literature have exactly characterized the secret message capacity for the case of a
single link \cite{czap2011secret}, a V-network (with 2 links) \cite{czap2013exploiting},
and a triangle network (with 3-links) \cite{czap2014triangle}, when only one of these channels
is eavesdropped; in all cases these are at most two-hop networks. Our work builds on these
results and further develops new achievability techniques and outer bounds for the case of
a multi-hop line network. The work in \cite{diggavi2013secure} has developed achievability
schemes and bounds for arbitrary networks with erasures and feedback but not exact characterizations.
The work in \cite{cai2002secure} looks at error free networks, while the work in \cite{mills2008secure}
does not consider feedback; additionally, all these works do not allow intermediate nodes to
generate (possibly limited) randomness, as we do.

To develop our results, we introduce new achievability schemes and outer bounds. In our schemes,
we consider a spectrum of choices for the intermediate node (relay) private randomness, ranging
from the extreme case where each relay can generate unlimited private randomness, to the other
extreme case where each relay can generate no private randomness, and including all the cases
in-between (limited randomness). We provide an outer bound in the form of a Linear Program
(LP), that applies for arbitrary values of V, and uses a new technique to incorporate the available
randomness at the network nodes for the derivation of the outer bound constraints. We also
provide achievability algorithms for the cases V = 1 and V = N, that come from the solution
of an achievability LP, and employ new techniques to generate secret keys between the network
nodes. These algorithms use the available randomness at each node efficiently and illustrate
the dependency between the amount of available randomness and the achievable secret-message
rates. We prove that for V = 1 and V = N the outer bound LPs matches the achievability LPs,
and thus we have an exact characterization, that applies for all cases of private randomness
at the relays (unlimited, limited, no private randomness).

As a side result, we provide the exact secret key and secret message capacity characterization
of a source that has limited private randomness, and is connected to a destination over a single
erasure channel with feedback (eavesdropped by Eve). This is a generalization of the scenario
examined in \cite{czap2011secret}, which assumes unlimited private randomness at the source.
We use this result as a building block for characterizing the secrecy capacity over line networks;
indeed, intermediate relays, if they cannot have access to an unlimited private randomness
source, they are necessarily limited to use the randomness received by their predecessor node
in the line network.

Our results enable to make several interesting observations. First, we verify the usefulness
of erasures as well as feedback in securely sending messages and generating secret keys over
line networks. Indeed, assuming perfect channels in a line network gives a zero secret message
rate even in the case when $V=1$. Second, our results imply that having feedback between all
network nodes is unnecessary: we can achieve the same rates even if we only have feedback from
a node to its predecessor. Third, it is also interesting that designing an optimal achievability
scheme when intermediate nodes have private randomness is a polynomial-time problem over a
line network, while it is known that this a NP-hard problem over arbitrary networks \cite{huang2013secure}.
Finally, like in all previous cases for smaller networks \cite{czap2011secret}, a 2-phase
scheme where we generate secret keys and then consume them for message encryption (in each
hop) remains optimal.

The rest of this paper is organized as follows. Section~\ref{sec:System-model-and} describes
the system model; Section~\ref{sec:Main-Results} summarizes the main results of our work;
Section~\ref{sec:Broadcast-erasure-channel} presents the secret key and secret message achievability
algorithms of the broadcast erasure channel with feedback and limited randomness, while Sections~\ref{sec:One-Eve-line-network}
and \ref{sec:All-Eves-line-network} provide the line network achievability algorithms for
 $V=1$ and $V=N$, respectively. All outer bounds are delegate to the Appendices.

\section{\label{sec:System-model-and}System model and Notation}

\begin{figure*}
\begin{centering}
\includegraphics[scale=0.6]{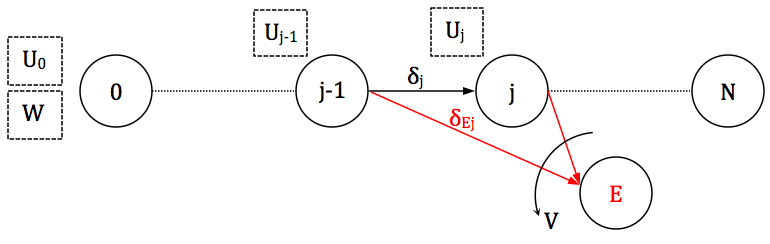}
\par\end{centering}

$\vspace{-0.3cm}$

\caption{\label{fig:Line}The general V-Eves Line Network. A subset with cardinality $V$ of the channels
is eavesdropped.}

$\vspace{-0.6cm}$
\end{figure*}

We consider a line network with $N$ hops, i.e. a network where the nodes are ordered and each
communicates through a channel with the next one, as shown in Fig.~\ref{fig:Line}. In our
case: 1) Each hop is a discrete memoryless broadcast erasure channel with two receivers: the
next node and potentially a passive eavesdropper (Eve). The broadcast channel is conditionally
independent (defined formally in the next paragraph) 2) We have is public state feedback. That
is, each node sends an ACK (or NACK) so that all other nodes (including Eve) learn whether
the packet transmission was successful. The source (node 0, or Alice) aims to end a message
W securely (formally defined later) to the destination (node N, or Bob).

We denote with $\mathcal{N}$ the set $[1,\ldots,N]$. We denote with $\mathcal{N}_{E}$ the
set of eavesdropped edges and with $V$ its cardinality. The notation $\mathcal{N}_{E}\underset{V}{\subset}\mathcal{N}$
is used to denote that $\mathcal{N}_{E}$ is a subset of $\mathcal{N}$ of cardinality $V$.
For a set $\mathcal{B}$ we define $\mathcal{B}_{-j}\triangleq\mathcal{B\backslash}\{j\}$.
Also, we denote $[j]\triangleq[1,...,j]$. We use $i$ as a time variable and $j$ to index
the nodes and the edges. Node $j$ is connected with node $j+1$ through edge (channel) $j$.
We denote with $W$ the message that has to be transmitted securely from node $0$ to node
$N$. The input to channel $j$ sent by node $j-1$ at time slot $i$ is denoted $X_{ji}$
and it is a length $L$ vector over $\mathbb{F}_{q}$. In the achievability algorithms we use
the convention that $L\log(q)=1$. We denote with $Y_{ji}$ and $Z_{ji}$ the $ith$ output
of the $jth$ channel, i.e., the vectors received by node $j$ and Eve respectively. We use
$\oslash$ as the symbol of an erasure. Channels are memoryless and conditionally independent,
i.e., $\Pr\{Y_{ji}^{n},Z_{ji}^{n}|X_{ji}^{n}\}=\underset{i=1}{\overset{n}{\prod}}\Pr\{Y_{ji}|X_{ji}\}\Pr\{Z_{ji}|X_{ji}\}$,
and

\begin{align*}
\Pr\{Y_{ji}|X_{ji}\} & =\begin{cases}
1-\delta_{j}, & Y_{ji}=X_{ji}\\
\delta_{j}, & Y_{ji}=\oslash,
\end{cases}
\end{align*}

\begin{eqnarray*}
\Pr\{Z_{ji}|X_{ji}\} & =\begin{cases}
1-\delta_{jE}, & Z_{ji}=X_{ji}\\
\delta_{jE}, & Z_{ji}=\oslash.
\end{cases}
\end{eqnarray*}
%Further, each channel operates independently. The probabilities $\delta_{j}$, $\delta_{jE}$
%are known by all parties.

Let $S_{ji}$ denote the random variable that describes the state of node $j$'s channel at
the $i$th transmission. $S_{ji}$ is a random variable with values in $\{B_{j},\emptyset\}$,
where $Pr\{S_{ji}=B_{j}\}=1-\delta_{j}$ meaning node $j$ correctly received the $i$th packet.
$S_{ji}$ is independent of $(X^{i},Y^{i-1},Z^{i-1},W)$. We model the feedback channel as
our nodes and Eve having access causally to the channel states, i.e.~before the $i$th transmission
they both know the vector $S^{i-1}$ (defined next). The notation $X_{j}^{i}$ is used to denote
the vector $(X_{j1},X_{j2},\dots,X_{ji})$, $X_{i}$ is used to denote the vector $(X_{1i},X_{2i},\dots,X_{Ni})$,
$X^{i}$ is used to denote the vector $(X_{1}^{i},X_{2}^{i},\dots,X_{N}^{i})$, $X_{[j]}^{i}$
is used to denote the vector $(X_{1}^{i},X_{2}^{i},\dots,X_{j}^{i})$ and similarly for $Y$,
$Z$, $S$. Furthermore, each node has access to a rate-limited private random source. We denote
by $U_{j}$ the available private random source at node $j$.

We will call the case where $V=1$ the One-Eve line network, the case where $V=N$, i.e., all
channels are eavesdropped, the All-Eves line network, and in-between cases the V-Eves line
network. 
\begin{defn}
\label{Def1}We say that $R_{SM}$ is an achievable secret message rate if for any $\epsilon>0$
and sufficiently large $n$ the following conditions hold for some functions $f_{ji,n}(\cdot),W_{B,n}(\cdot)$:
\begin{align}
X_{ji}=\begin{cases}
f_{ji,n}(W,U_{0},S^{i-1}) & \; j=1\\
f_{ji,n}(Y_{j-1}^{i-1},U_{j-1},S^{i-1}) & \; j=2,\ldots,N
\end{cases}\label{eq:def1_1}
\end{align}
where the message $W$ is uniformly distributed over $\{1,2,\ldots,2^{n(R_{SM}-\epsilon)}\}$.
Node $N$ is able to recover $W$ with high probability: 
\begin{align}
\hat{W}=W_{B,n}(Y_{N}^{n}),\label{eq:def1_2}\\
\Pr\{\hat{W}\neq W\}<\epsilon.\label{eq:def1_3}
\end{align}
Eve gains negligible useful information: 
\begin{align}
\begin{cases}
\textrm{One-Eve line network: } & I(W;Z_{j}^{n}S^{n})<\epsilon\:\forall j\in\mathcal{N}\\
\textrm{All-Eves line network: } & I(W;Z^{n}S^{n})<\epsilon\\
\textrm{V-Eves line network: } & I(W;Z_{\mathcal{N}_{E}}^{n}S^{n})<\epsilon\:\forall\mathcal{N_{\textrm{E}}}\underset{V}{\subset}\mathcal{N}
\end{cases}\label{eq:def1_5}
\end{align}
The supremum of all achievable secret message rates is the secret message capacity of the network
denoted by $C_{SM}$. 
\end{defn}
\begin{figure}
\centering{}\includegraphics[scale=0.6]{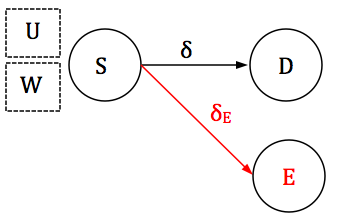} \caption{\label{fig:1-hop-Network}Broadcast erasure channel with feedback and limited randomness}
\end{figure}

\begin{defn}
We say that $R_{SK}$ is an achievable secret key rate if for any $\epsilon>0$ and sufficiently
large $n$ the following conditions hold. For a function $f$ node $1$ creates, 
\[
K=f(X_{1}^{n},S^{n},U_{0}),
\]
where $K$ is the random variable representing the key which takes values in the set $\mathcal{K}$
%that has cardinality $R_{SK}=\left|\mathcal{K}\right|$. 
Node $N$ creates, 
\[
\hat{K}=f(Y_{N}^{n},S^{n}),
\]
which also takes values in $\mathcal{K}$. The same key is computed with high probability:
\begin{equation}
Pr\{K\neq\hat{K}\}<\epsilon.\label{eq:ErrorZero}
\end{equation}
The key is (almost) uniform: 
\begin{equation}
H(K)>\log\left|\mathcal{K}\right|-\epsilon.\label{eq:Uniformity}
\end{equation}
The key remains secret form Eve: 
\begin{equation}
\begin{cases}
\textrm{One-Eve line network: } & I(K;Z_{j}^{n}S^{n})<\epsilon,\quad\forall j\in\mathcal{N}\\
\textrm{All-Eves line network: } & I(K;Z^{n}S^{n})<\epsilon\\
\textrm{V-Eves line network: } & I(K;Z_{\mathcal{N}_{E}}^{n}S^{n})<\epsilon,\quad\forall\mathcal{N_{\textrm{E}}}\underset{V}{\subset}\mathcal{N}
\end{cases}\label{eq:Security}
\end{equation}
The supremum of all achievable secret key rates is the secret key capacity of the network denoted
by $C_{SK}$. 

$\vspace{-0.8cm}$
\end{defn}

\section{\label{sec:Main-Results}Main Results}

%========================================
We here collect the main results of our work. $\vspace{-0.5cm}$

\subsection{Broadcast channel with limited randomness at the source}
\begin{thm}
\label{thm:The-secret-key}The secret key capacity $C_{SK}$ of the broadcast erasure channel
with state feedback and a limited randomness source $U$ with $ $$D\triangleq\underset{n\rightarrow\infty}{\liminf}\frac{H\left(U\right)}{n}$
equals: 
\[
C_{SK}=\min\left\{ \frac{D\delta_{E}(1-\delta)}{1-\delta\delta_{E}},(1-\delta)\delta_{E}\right\} .
\]
\label{thm:The-secret-message}
\end{thm}
$\vspace{-0.5cm}$

$\vspace{-1cm}$
\begin{thm}
The secret message capacity $C_{SM}$ of the broadcast erasure channel with a limited randomness
source $U$ with $ $$D\triangleq\underset{n\rightarrow\infty}{\liminf}\frac{H\left(U\right)}{n}$
equals:

\[
C_{SM}=\min\left\{ \frac{\left(1-\delta\right)\delta_{E}}{1-\delta_{E}}D,\left(1-\delta\right)\delta_{E}\frac{1-\delta\delta_{E}}{1-\delta\delta_{E}^{2}}\right\} .
\]

\end{thm}
This theorem generalizes the result of \cite{czap2011secret} and shows how secrecy depends
on the available randomness $D$. The achievability scheme is presented in Section~\ref{sec:Broadcast-erasure-channel}
and the converse proof in Appendix A.

\subsection{One-Eve Line Network}

%The next two theorems characterize
%the secret message capacity of the One-Eve in the case that the intermediate nodes cannot create
%their own randomness and in the case that they have access to random sources with limited rate.

\begin{thm}
\label{thm:The-secret-message-1-1}The secret message capacity of the One-Eve line network,
with erasures, state feedback, no private randomness at the intermediate nodes and unlimited
private randomness at the source, equals the solution of the following LP: 
\begin{eqnarray*}
max & m,\\
s.t.\:\ensuremath{\forall j\in\mathcal{N}}:\\
\frac{1-\delta_{jE}}{1-\delta_{j}\delta_{jE}}m & \leq & k_{j}\\
\frac{k_{j}}{(1-\delta_{j})\delta_{jE}}+\frac{m}{1-\delta_{j}} & \leq & 1\\
k_{j} & \leq & d_{j-1}\frac{\delta_{jE}(1-\delta_{j})}{1-\delta_{j}\delta_{jE}}\\
d_{j}+m & \leq & 1-\delta_{j}\\
d_{j} & \leq & d_{j-1},\: j>1\\
k_{j},d_{j},m & \geq & 0
\end{eqnarray*}

\end{thm}

\begin{thm}
\label{thm:The-secret -mess-gen-1}The secret message capacity $C_{SK}$ of the One-Eve line
network with erasures, state feedback, and at intermediate nodes limited randomness sources
$U_{j}$, $\forall j\in\mathcal{N}$ with $ $$D_{j}\triangleq\underset{n\rightarrow\infty}{\liminf}\frac{H\left(U_{j}\right)}{n}$,
equals the solution of the following LP: 
\begin{eqnarray*}
max & m,\\
s.t.\:\ensuremath{\forall j\in\mathcal{N}}:\\
\frac{1-\delta_{jE}}{1-\delta_{j}\delta_{jE}}m & \leq & k_{j}\\
\frac{k_{j}}{(1-\delta_{j})\delta_{jE}}+\frac{m}{1-\delta_{j}} & \leq & 1\\
k_{j} & \leq & (d_{j-1}+D_{j-1})\frac{\delta_{jE}(1-\delta_{j})}{1-\delta_{j}\delta_{jE}}\\
d_{j}+m & \leq & 1-\delta_{j}\\
d_{j} & \leq & d_{j-1}+D_{j-1}\\
k_{j},d_{j},m & \geq & 0,
\end{eqnarray*}
where $d_{0}\triangleq0$. 
\end{thm}
The achievability scheme and the LP variables are explained in Sec.~\ref{sec:One-Eve-line-network}
; the outer bound is in Appendix B.$\vspace{-0.3cm}$

\subsection{All-Eves Line Network}
\begin{thm}
\label{thm:The-secret-message-1-1-1}The secret message capacity of the erasure All-Eves line
network, with erasures, state feedback, no randomness at the intermediate nodes and unlimited
randomness at the source, is the solution of the following LP: 
\end{thm}
\[
\begin{array}{ccc}
max & m\\
s.t.\:\ensuremath{\forall j\in\mathcal{N}}:\\
\frac{1-\delta_{jE}}{1-\delta_{j}\delta_{jE}}m & \leq & k_{j}-d_{j}\\
\frac{k_{j}}{(1-\delta_{j})\delta_{jE}}+\frac{m}{1-\delta_{j}} & \leq & 1\\
k_{j} & \leq & d_{j-1}\frac{\delta_{jE}(1-\delta_{j})}{1-\delta_{j}\delta_{jE}},\: j>1\\
k_{j},m & \geq & 0
\end{array}
\]

\begin{thm}
\label{thm:The-secret -mess-gen-1-1}The secret message capacity $C_{SK}$ of the All-Eves
line network with erasures, state feedback, and limited randomness sources $U_{j}$, $\forall j\in\mathcal{N}$
with $ $$D_{j}\triangleq\underset{n\rightarrow\infty}{\liminf}\frac{H\left(U_{j}\right)}{n}$
at intermediate nodes, is the solution of the following LP: 
\[
\begin{array}{ccc}
max & m\\
s.t.\ensuremath{\:\forall j\in\mathcal{N}}:\\
\frac{1-\delta_{jE}}{1-\delta_{j}\delta_{jE}}m & \leq & k_{j}-d_{j}\\
\frac{k_{j}}{(1-\delta_{j})\delta_{jE}}+\frac{m}{1-\delta_{j}} & \leq & 1\\
k_{j} & \leq & \left(d_{j-1}+D_{j-1}\right)\frac{\delta_{jE}(1-\delta_{j})}{1-\delta_{j}\delta_{jE}}\\
k_{j},m & \geq & 0,
\end{array}
\]
where $d_{0}\triangleq0$. 
\end{thm}
The achievability scheme and the LP variables are explained in Sec.~\ref{sec:All-Eves-line-network}
; the outer bound is in Appendix B.

\subsection{Outer bound for the V-Eves network}

The following outer bound, provided in Appendix B, applies for all V.
\begin{thm}
\label{thm:OuterBound}The secret message capacity $C_{SK}$ of the V-Eves line network with
erasures, state feedback, and limited randomness sources $U_{j}$, $\forall j\in\mathcal{N}$
with $ $$D_{j}\triangleq\underset{n\rightarrow\infty}{\liminf}\frac{H\left(U_{j}\right)}{n}$
is smaller or equal to the solution of the following LP, 
\[
\begin{array}{ccc}
\\
max & m\\
s.t.\:\forall\mathcal{N_{\textrm{E}}}\underset{V}{\subset}\mathcal{N}:\\
\forall j\in\mathcal{N_{\textrm{E}}}:\\
\frac{1-\delta_{jE}}{1-\delta_{j}\delta_{jE}}m & \leq & k_{j}^{\mathcal{N_{\textrm{E}}}}-d_{j}^{\mathcal{N_{\textrm{E}}}}\\
\frac{k_{j}^{\mathcal{N_{\textrm{E}}}}}{(1-\delta_{j})\delta_{jE}}+\frac{m}{1-\delta_{j}} & \leq & 1\\
k_{j}^{\mathcal{N_{\textrm{E}}}} & \leq & \left(d_{j-1}^{\mathcal{N_{\textrm{E}}}}+D_{j-1}\right)\frac{\delta_{jE}(1-\delta_{j})}{1-\delta_{j}\delta_{jE}}\\
\forall j\in\mathcal{N}-\mathcal{N_{\textrm{E}}}:\\
d_{j}^{\mathcal{N_{\textrm{E}}}}+m & \leq & 1-\delta_{j}\\
d_{j}^{\mathcal{N_{\textrm{E}}}} & \leq & d_{j-1}^{\mathcal{N_{\textrm{E}}}}+D_{j-1},\: j>1\\
k_{j}^{\mathcal{N_{\textrm{E}}}},d_{j}^{\mathcal{N_{\textrm{E}}}},m & \geq & 0
\end{array}
\]
where $d_{0}^{\mathcal{N_{\textrm{E}}}}\triangleq0$ $\,\forall\mathcal{N_{\textrm{E}}}\underset{V}{\subset}\mathcal{N}$. 
\end{thm}
In order to derive this outer bound we developed a number of techniques that may be useful
for other networks. First we considered all the different $\left(\begin{array}{c}
N\\
V
\end{array}\right)$ ``positionings'' of Eve in the channels and we derived constraints for all these cases.
Next, in order to connect the constraints for each channel in the line network, we identified
the information theoretic term that plays the role of ``available randomness'' for the secret
key generation in the next channel. Since we want this randomness to be unknown by Eve, this
term has to represent the ``secure available randomness''. Putting all these constraints,
for each channel and for each ``positioning'' of Eve, together, results in the provided outer
bound.

\section{\label{sec:Broadcast-erasure-channel} Single channel with limited randomness at the source:
a building block for line networks}

%=============================================================================
In this Section we present the achievability algorithm for secret key generation and secret
message transmission of the broadcast erasure channel with feedback and limited randomness
depicted in Fig. \ref{fig:1-hop-Network}. This serves as a building block of our line network
algorithms: indeed, each edge from node $j$ to node $j+1$ in the line network can be viewed
as a broadcast channel with potentially limited randomness at the source.

From previous work \cite{czap2011secret}, we know that, when the source in Fig. \ref{fig:1-hop-Network}
has unlimited randomness, the optimal achievability scheme involves two stages: in the first
(key generation phase), the source sends at each transmission a different random packet so
as to create a secret key with the destination; in the second (message transmission phase),
the source uses the secret key to securely send the message.

When the source has limited randomness, we prove in this paper that the optimal scheme is still
a two-phase scheme, where again in the first phase we generate a secret key, and in the second
phase we use the key to secure the message. What changes from the unlimited randomness case,
is how we generate the secret key in the first phase, i.e., how do we best use the limited
randomness at the source so as to create a maximum rate key between the source and the destination.
Additionally, because we want to use this scheme as a building block for the line network,
we are interested in a second goal as well: we want the destination to receive as many random
packets from the source as possible (independently of whether Eve has overhead these packets
or not). The reason for this is that, if intermediate nodes in a line network do not generate
(enough) private randomness, they need to rely on the randomness the receive from previous
nodes (that is, node j+1 relies on node j to receive random packets); thus we want to maximize
the amount of random packets they receive. In summary, we set two goals: 
\begin{itemize}
\item {\bf G1:} Given limited randomness at the source, achieve the optimal key-generation rate. 
\item {\bf G2:} Given limited randomness at the source, and optimal key-generation rate, maximize
the amount of randomness that the receiver gets from the source. 
\end{itemize}

\subsection{Schemes that achieve Goal 1 (G1)}

Table \ref{tab:Eff} compares three algorithms for using the source randomness (assuming rate
D for the source) for key generation: 
\begin{enumerate}
\item \textbf{KG} sends a different random packet at each transmission. 
\item \textbf{ARQ} repeats each random packet until the destination receives it. 
\item \textbf{MDS-exp} expands the $D$ random packets by multiplying them with an MDS matrix of
size $nD\times\frac{nD}{1-\delta\delta_{E}}$ and transmits each of the resulting packets once. 
\end{enumerate}
Each of these schemes can be optimal wrt our first goal (max key rate) in different scenaria.
The first scheme (KG) is optimal when $D\geq1$ (we have a new random packet to send at every
transmission). A main property it ensures is that, all packets that Eve receives and the destination
does not, will not be useful to Eve, as they will not be used for the key generation. However,
it is inefficient in ensuring this property, because, there will exist random packet transmissions
that neither Eve nor the destination will receive; and thus these random packets will be \textquotedbl{}wasted\textquotedbl{}.
The second method (ARQ) ensures that every random packet does reach the destination. In this
case we do not waste any random packets, but since each packet is transmitted multiple times,
Eve will observe it with higher probability. This scheme is optimal only when the source randomness
is lower than $1-\delta$, i.e. we have enough time to send all the random packets we have
with ARQ. The third method (MDS-exp) achieves the same property as the first, i.e., packets
Eve receives and the receiver does not, are not useful to Eve, but avoids the inefficiency
in the random packet consumption by expanding in advance the random keys. This scheme is optimal
in key generation for the general case of limited randomness at the source. The next two theorems
prove that ARQ and MDS-exp algorithms preserve the security condition \ref{eq:Security}.

\begin{table}
\begin{centering}
\caption{\label{tab:Eff} Comparing three ways to use the source randomness for secret key generation. }

\par\end{centering}

\centering{}%
\begin{tabular}{|c|c|c|c|}
\hline 
 & KG  & ARQ  & MDS-exp\tabularnewline
\hline 
\hline 
Keys/transmission  & $\delta_{E}(1-\delta)$  & $\frac{\delta_{E}(1-\delta)^{2}}{1-\delta\delta_{E}}$  & $\delta_{E}(1-\delta)$\tabularnewline
\hline 
Consumed  & \multirow{2}{*}{$1$ } & \multirow{2}{*}{$1-\delta\delta_{E}$ } & \multirow{2}{*}{$1-\delta\delta_{E}$}\tabularnewline
Randomness/transmission &  &  & \tabularnewline
\hline 
\end{tabular}
\end{table}

\begin{thm}
\label{thm:The-algorithm-that}The algorithm that achieves the secret key capacity in the case
that the available randomness is $D\leq1-\delta$, involves the transmission of the $D$ packets
with ARQ. After the transmission the receiver creates linear combinations of rate $\frac{(1-\delta)\delta_{E}D}{1-\delta\delta_{E}}$$ $,
whose coefficients are determined by the rows of an MDS matrix of size $ $$\frac{(1-\delta)\delta_{E}nD}{1-\delta\delta_{E}}\times nD$.
These $\frac{(1-\delta)\delta_{E}D}{1-\delta\delta_{E}}$ key packets preserve the security
condition (\ref{eq:Security}).\end{thm}
\begin{IEEEproof}
The achievability part is an application of Lemma 1 of \cite{czap2013exploiting} for $r_{1}=D$.
The converse is proved for a more general case in \ref{thm:(Coverse of SK)} in appendix A. \end{IEEEproof}
\begin{thm}
\label{thm:The-following-algorithm}The following algorithm creates a secret key that preserves
the security condition (\ref{eq:Security}). The transmitter multiplies the $D$ random packets
with an MDS matrix of size $nD\times\frac{nD}{1-\delta\delta_{E}}$. Then these $\frac{nD}{1-\delta\delta_{E}}$
packets are transmitted once and the receiver creates linear combinations of rate $\frac{(1-\delta)\delta_{E}D}{1-\delta\delta_{E}}$$ $,
whose coefficients are determined by the rows of an MDS matrix of size $ $$\frac{(1-\delta)\delta_{E}D}{1-\delta\delta_{E}}\times\frac{nD}{1-\delta\delta_{E}}$.\end{thm}
\begin{IEEEproof}
It is an application of Lemma 3 of \cite{czap2013exploiting} for $c_{2}=0$ and $c=D$. 
\end{IEEEproof}
$\vspace{-0.5cm}$

\subsection{A scheme that optimizes Goal 1 (G1) and Goal 2 (G2)}

\begin{figure}
\centering{}$\;$ \includegraphics[scale=0.6]{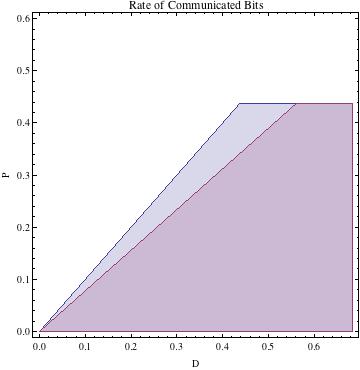}\caption{\label{fig:-Comparison-of-P}Comparison of communicated bit rate (to the next node) of MDS-exp
(purple region) with MDS-exp/ARQ (blue region).}
\end{figure}

%========================================================
Although the MDS-exp is optimal in terms of key-generation, it turns out that when we want
to also optimize our second goal (convey maximum randomness to the destination), the optimal
scheme timeshares between MDS-exp and ARQ. The intuition is the following. The MDS expansion
and the ARQ schemes, both using the same amount of randomness $D$, could create the same amount
of secret key $\frac{nD\delta_{E}(1-\delta)}{1-\delta\delta_{E}}$. However, they do not have
the same time efficiency, since the ARQ scheme would use more time slots, transmitting more
packets: $\frac{nD}{1-\delta}$ compared to $\frac{nD}{1-\delta\delta_{E}}$. So, doing time
sharing between these two schemes can both create the maximum secret key and communicate the
maximum amount of packets to the receiver. Table~\ref{tab:Summary} and Fig. \ref{fig:-Comparison-of-P}
show the difference in the secret key rate and the communicated packet rate between these algorithms.
We next briefly analyze the MDS-exp/ARQ algorithm.

\begin{table*}
\begin{centering}
\caption{\label{tab:Summary} Comparison of schemes wrt to G1 and G2 }

\par\end{centering}

\centering{}%
\begin{tabular}{|c|c|c|}
\hline 
 & Secret key rate (G1)  & Randomness communicated to the next node (G2) \tabularnewline
\hline 
\hline 
ARQ  & $R_{SK}=\begin{cases}
D\frac{(1-\delta)\delta_{E}}{1-\delta\delta_{E}} & \; if\: D<1-\delta\\
\frac{(1-\delta)^{2}\delta_{E}}{1-\delta\delta_{E}} & \; if\: D>1-\delta
\end{cases}$  & $P=\begin{cases}
D & \; if\: D<1-\delta\\
1-\delta & \; if\: D\geq1-\delta
\end{cases}$\tabularnewline
\hline 
MDS-exp  & $R_{SK}=\begin{cases}
\frac{D\delta_{E}(1-\delta)}{1-\delta\delta_{E}} & \; if\: D<{1-\delta\delta_{E}}\\
(1-\delta)\delta_{E} & \; if\: D\geq{1-\delta\delta_{E}}
\end{cases}$  & $P=\begin{cases}
\frac{D(1-\delta)}{1-\delta\delta_{E}} & \; if\: D<{1-\delta\delta_{E}}\\
1-\delta & \; if\: D\geq{1-\delta\delta_{E}}
\end{cases}$\tabularnewline
\hline 
MDS-exp/ARQ  & $R_{SK}=\begin{cases}
\frac{D\delta_{E}(1-\delta)}{1-\delta\delta_{E}} & \; if\: D<1-\delta\delta_{E}\\
(1-\delta)\delta_{E} & \; if\: D\geq1-\delta\delta_{E}
\end{cases}$  & $P=\begin{cases}
D & \; if\: D<1-\delta\\
1-\delta & \; if\: D\geq1-\delta
\end{cases}$\tabularnewline
\hline 
\end{tabular}
\end{table*}

\subsubsection{Analysis of the algorithm}

We chose a parameter $a$ for time-sharing so that we use all available time: 
\[
\frac{aD}{1-\delta\delta_{E}}+\frac{(1-a)D}{1-\delta}=1.
\]
Calculating: 
\[
aD(1-\delta)+(1-a)D(1-\delta\delta_{E})=(1-\delta)(1-\delta\delta_{E})\Rightarrow
\]

\[
a=\frac{D(1-\delta\delta_{E})-(1-\delta)(1-\delta\delta_{E})}{D\delta(1-\delta_{E})},
\]

\begin{itemize}
\item When $a<0$, we do ARQ only, since 
\[
\frac{D(1-\delta\delta_{E})-(1-\delta)(1-\delta\delta_{E})}{D\delta(1-\delta_{E})}<0\Rightarrow D<1-\delta
\]
Thus, we have enough time to do ARQ for all the packets.

\begin{itemize}
\item The key we create is, 
\[
R_{SK}=\frac{D(1-\delta)\delta_{E}}{1-\delta\delta_{E}},
\]
which is the optimal. 
\item The packets communicated have a rate of, 
\[
P=D,
\]
which is the maximum. 
\end{itemize}
\item When $a>1$, we do MDS expansion only, since 
\[
\frac{D(1-\delta\delta_{E})-(1-\delta)(1-\delta\delta_{E})}{D\delta(1-\delta_{E})}>1\Rightarrow D>1-\delta\delta_{E}
\]
Thus, we have enough randomness to do MDS expansion only.

\begin{itemize}
\item The key we create is, 
\[
R_{SK}=(1-\delta)\delta_{E},
\]
which is the optimal. 
\item The packets communicated have a rate of, 
\[
P=1-\delta,
\]
which is the maximum. 
\end{itemize}
\item When $0\leq a\leq1$, we do MDS expansion for $a$ percent of the packets and ARQ for the rest.

\begin{itemize}
\item The key we create is,
\begin{align*}
R_{SK} & =\frac{aD}{1-\delta\delta_{E}}(1-\delta)\delta_{E}+(1-a)D\frac{(1-\delta)\delta_{E}}{1-\delta\delta_{E}}\\
 & =\frac{D\delta_{E}(1-\delta)}{1-\delta\delta_{E}}
\end{align*}
which is the optimal. 
\item The packets communicated have a rate of, 
\[
P=1-\delta,
\]
since we send an innovative (for the next node) packet in each time slot, and is the maximum. 
\end{itemize}
\item Summing up:

\begin{itemize}
\item The algorithm creates, 
\[
R_{SK}=\begin{cases}
\frac{D\delta_{E}(1-\delta)}{1-\delta\delta_{E}} & \; if\: D<1-\delta\delta_{E}\\
(1-\delta)\delta_{E} & \; if\: D\geq1-\delta\delta_{E}
\end{cases},
\]
which is the optimal. 
\item The packets communicated have a rate of, 
\[
P=\begin{cases}
D & \; if\: D<1-\delta\\
1-\delta & \; if\: D\geq1-\delta
\end{cases},
\]
which is the maximum. 
\end{itemize}
\end{itemize}
Table \ref{tab:Summary} summarizes these results. The secrecy of this algorithm depends on
the secrecy of the ARQ and MDS expansion phases, which are secure by Theorems \ref{thm:The-algorithm-that}
and \ref{thm:The-following-algorithm}.

\subsubsection{Expression through LP of the secret key capacity}

Although we have exact characterizations in this case, it is interesting to note that the optimal
solutions can be expressed through LP formulations. The secret key capacity can be expressed
as the solution of the following LP: 
\begin{eqnarray*}
max & k\\
\frac{k}{(1-\delta)\delta_{E}} & \leq & 1\\
k & \leq & D\frac{\delta_{E}(1-\delta)}{1-\delta\delta_{E}}\\
k & \geq & 0
\end{eqnarray*}
Solving this LP gives us the rate depicted in Table \ref{tab:Summary}.

\subsubsection{Expression through LP of the secret message capacity}

The secret message capacity can be expressed as the solution of the following LP: 
\begin{eqnarray*}
max & m\\
\frac{1-\delta_{E}}{1-\delta\delta_{E}}m & \leq & k\\
\frac{k}{(1-\delta)\delta_{E}}+\frac{m}{1-\delta} & \leq & 1\\
k & \leq & D\frac{\delta_{E}(1-\delta)}{1-\delta\delta_{E}}\\
k,m & \geq & 0
\end{eqnarray*}
The variables $m$, $k$ and $D$ represent the message rate, the key that we create and the
available randomness, respectively. The first inequality is a security constraint. The key
that is consumed has to be smaller than the one we created. The second inequality is a time
constraint. The length of the key generation phase plus the length of the message sending phase
have to not exceed the available time. These two inequalities alone describe the algorithm
in \cite{czap2011secret}. The third is the constraint imposed on the secret key that we can
create due to the limited available randomness. All converse proofs are delegated to Appendix
A.

\section{\label{sec:One-Eve-line-network}One-Eve line network}

%=================================================

We here consider the case where Eve eavesdrops a single channel. During the key generation
phase, each node $j$ creates a key $K_{j-1}$ with node $j-1$ and a key $K_{j}$ node $j+1$.
That is, we always create one-hop keys. During the message transmission phase, node $j$ receives
the messages encrypted with $K_{j-1}$; it decrypts it, re-encrypts it with key $K_{j}$ and
proceeds to send it to node $j+1$. Depending on how much randomness the intermediate nodes
have, we create the keys in different ways as described next. 
\begin{itemize}
\item {\em Unlimited private randomness at intermediate nodes}: node $j$ creates the key $K_{j}$
using only its own private randomness. 
\item {\em No private randomness at intermediate nodes}: node $j$ creates the key $K_{j}$ using
the random packets it has received from node $j-1$. 
\item {\em Limited randomness at intermediate nodes}: node $j$ uses both its own private randomness
as well the random packets it has received from node $j-1$. 
\end{itemize}
When there is no or limited private randomness in intermediate nodes, each node $j$ uses the
scheme we described in Theorem~\ref{thm:The-following-algorithm} to create the next hop key.
Note that because Eve is present in only one channel, all random packets that node $j$ receives
can be used to create the next hop key (if Eve is in the next hop, she has not received these
packets). Thus Theorem \ref{thm:The-following-algorithm} applies again for each hop. This
proves that the algorithm preserves the security requirement \ref{eq:def1_5}. We next briefly
describe the LPs that achieve the optimal solution (the matching outer bound is provided Appendix
B).

Theorem~\ref{thm:The-secret-message-1-1} characterizes the secret message capacity when there
is no private randomness. The variables $m$, $k_{j}$ and $d_{j}$ represent the message rate,
the key that we create at hop $j$ and the available randomness at node $j$, respectively.
The first three inequalities (for each hop $j$) are the constraints of the Broadcast erasure
channel with feedback and limited randomness. The first inequality is a security constraint.
The key that is consumed has to be smaller than the one we created. The second inequality is
a time constraint. The length of the key generation phase plus the length of the message sending
phase have to not exceed the available time. The third is the constraint imposed on the secret
key that we can create at hop $j$ due to the fact that node $j-1$ has only $d_{j-1}$ available
randomness. The last two inequalities describe the flow of randomness. The random packets that
we can send to the next node are smaller or equal to the ones we have, and smaller or equal
to the ones that we can send in the available time. This is exactly what the MDS-exp/ARQ algorithm
achieves, as we can see in Table \ref{tab:Summary}. This completes the presentation of the
achievability algorithm of theorem \ref{thm:The-secret-message-1-1}.

Theorem \ref{thm:The-secret -mess-gen-1} is a direct generalization for the case that the
relay nodes have access to limited randomness sources. In this case the randomness available
in each node is the sum of the randomness received from the previous node plus the extra randomness
its random source produces. In appendix B the outer bound of the general case V-Eves channel
is proved and for $V=1$ matches this LP. It involves the construction of a converse LP equivalent
to this one, where each information term corresponds to each variable in this LP.

Finally, when each node has an unlimited randomness source, the LP consists only of the first
three inequalities (for each $j$). In this case the secret message capacity of the line network
can be interpreted as a cut-type result: it is the minimum secret key capacity of the hops
of the line network.

\paragraph*{Extensions}

Given the LPs we already have it is straightforward to create several extensions. For instance,
if there are constraints (say, there are $K$ of them) on limited randomness sources, i.e.,
$f_{k}\left(\left(D_{j}\right)_{j\in\mathcal{N}}\right)\leq0\;\forall k\in[K]$, the above
LP can be augmented with the these inequalities. As another example, if we want to minimize
the ``cost'' of the extra randomness sources (say we have a cost function $g$) for a specific
secret message rate $m^{'}\leq C_{SK}$, then we can use the LP:

\begin{eqnarray*}
\end{eqnarray*}
\begin{eqnarray*}
min &  & g\left(\left(D_{j}\right)_{j\in\mathcal{N}}\right),\\
s.t.\ensuremath{\:\forall j\in\mathcal{N}}:\\
\frac{1-\delta_{jE}}{1-\delta_{j}\delta_{jE}}m^{'} & \leq & k_{j}\\
\frac{k_{j}}{(1-\delta_{j})\delta_{jE}}+\frac{m^{'}}{1-\delta_{j}} & \leq & 1\\
k_{j} & \leq & (d_{j-1}+D_{j-1})\frac{\delta_{jE}(1-\delta_{j})}{1-\delta_{j}\delta_{jE}}\\
d_{j}+m^{'} & \leq & 1-\delta_{j}\\
d_{j} & \leq & d_{j-1}+D_{j-1}\\
\textrm{ and s.t. }\forall k\in\mathcal{K}:\\
f_{k}\left(\left(D_{j}\right)_{j\in\mathcal{N}}\right) & \leq & 0
\end{eqnarray*}

\section{\label{sec:All-Eves-line-network}All-Eves line network}

%===============================================

The main difference in the case where Eve is present in all edges is that, unlike the previous
$V=1$ case, randomness that has been used to create a key for a specific hop can not be used
to create keys for following hops. Clearly, when each node has unlimited private randomness,
all one hop keys are independent from each other, and thus the same scheme that works for $V=1$
also works for $V=N$. In the case of no (or limited) intermediate node randomness, when node
$j$ receives random packets from node $j-1$, it splits these packets into two parts: one
part is used to create the key $K_{j-1}$, and the other part is going to be forwarded towards
node $j$, to form the key $K_{j}$ as well as potentially subsequent channel keys. Since the
packets used to create the key $K_{j-1}$ are not forwarded towards node $j+1$, Theorem \ref{thm:The-following-algorithm}
applies again for each hop, and thus the security requirement \ref{eq:def1_5} is satisfied.
We let the linear program decide how to split the received randomness.

Theorem \ref{thm:The-secret-message-1-1-1} characterizes the secret message capacity when
there no private randomness at intermediate nodes; we next explain the variables in the LP.
The variables $m$, $k_{j}$ and $d_{j}$ represent the message rate, the key that we create
at hop $j$ and the available randomness at node $j$, respectively. The first inequality (for
each $j$) is a security constraint. The key that is consumed has to be smaller than the one
we created minus the packets we are going to use in the next channel. The second inequality
is a time constraint. The length of the key generation phase plus the length of the message
sending phase have to not exceed the available time. The third is the constraint imposed on
the secret key that we can create at hop $j$ due to the fact that node $j-1$ has only $d_{j-1}$
available randomness.

Theorem \ref{thm:The-secret -mess-gen-1-1} is a direct generalization for the case that the
relay nodes have access to limited randomness sources. In this case the secure randomness available
in each node is the sum of the key packets that were not consumed in the protection of the
message in the previous channel plus the extra randomness the node's random source produces.

\textit{Extensions:} Similarly to the $V=1$ case, we can extend the presented LPs for the
case where some nodes have constraints on the randomness they can generate, and for the case
where there is a cost associated generating source randomness. 

\textcolor{black}{\bibliographystyle{plain}
\bibliography{BECwithSI}
}

\appendices{}

\section*{Appendix A}

In this section we prove the converse of Theorem \ref{thm:The-secret-key} and Theorem \ref{thm:The-secret-message}.
Table \ref{tab:Summary-of-notation} summarizes the notation used in this paper.

\textcolor{black}{}
\begin{table}
\begin{centering}
\textcolor{black}{\caption{\textcolor{black}{\label{tab:Summary-of-notation}Summary of notation}}
\label{table:1}}
\par\end{centering}

\centering{}\textcolor{black}{}%
\begin{tabular}{cl}
\hline 
\textcolor{black}{$\boldsymbol{W}$} & \multirow{1}{*}{\textcolor{black}{The message.}}\tabularnewline
\textcolor{black}{$X_{ji}$} & \multirow{1}{*}{\textcolor{black}{The symbol sent by node $j-1$ in the $i$th time slot on channel $j$.}}\tabularnewline
\textcolor{black}{$Y_{ji}$} & \multirow{1}{*}{\textcolor{black}{The symbol received by node $j$ in the $i$th time slot on channel $j$.}}\tabularnewline
\textcolor{black}{$Z_{ji}$} & \multirow{1}{*}{\textcolor{black}{The symbol received by the eve in the $i$th time slot on channel $j$.}}\tabularnewline
\textcolor{black}{$S_{ji}$} & \multirow{1}{*}{\textcolor{black}{The public feedback at $i$th time slot on channel $j$.}}\tabularnewline
\textcolor{black}{$\boldsymbol{U}_{j}$} & \multicolumn{1}{l}{The limited randomness \textcolor{black}{source on channel $j$}.}\tabularnewline
\textcolor{black}{$R_{SM}$} & \textcolor{black}{The secret message rate.}\tabularnewline
$D_{j}$ & \textcolor{black}{The rate of th}e\textcolor{black}{{} limited randomness source on channel $j$}.\tabularnewline
$\mathbf{K}$ & The secret key.\tabularnewline
\textcolor{black}{$R_{SK}$} & \textcolor{black}{The secret key rate.}\tabularnewline
\hline 
\end{tabular}
\end{table}
The next general lemma is going to be used extensively in the proofs that follow. It is a generalization
of the corresponding lemma in \cite{czap2011secret}.
\begin{lem}
\label{lem:C}It is for $\mathcal{B}\subseteq[1,...,j]$:

\[
(1-\delta_{j})\delta_{jE}\sum_{i=1}^{n}H(X_{ji}|Y_{j}^{i-1}Z_{j}^{i-1}S_{j}^{i-1}WS_{\mathcal{N}_{-j}}^{n}Z_{\mathcal{B}_{-j}}^{n})-(1-\delta_{jE})\sum_{i=1}^{n}I(Y_{j}^{i-1}S_{j}^{i-1};X_{ji}|Z_{j}^{i-1}S_{j}^{i-1}WS_{\mathcal{N}_{-j}}^{n}Z_{\mathcal{B}_{-j}}^{n})\geq H\left(Y_{j}^{n}\mid WS_{\mathcal{N}}^{n}Z_{\mathcal{B}}^{n}\right)
\]
\end{lem}
\begin{IEEEproof}
\begin{align*}
 & H(Y_{j}^{n}|Z_{\mathcal{B}}^{n}S_{\mathcal{N}}^{n}W)=\\
={} & H(Y_{j}^{n}S_{j}^{n}|Z_{\mathcal{B}}^{n}S_{\mathcal{N}}^{n}W)\\
={} & H(Y_{j}^{n-1}S_{j}^{n-1}|Z_{\mathcal{B}}^{n}S^{n}W)+H(Y_{jn}S_{n}|Y_{j}^{n-1}Z_{\mathcal{B}}^{n}S^{n}W)\\
={} & H(Y_{j}^{n-1}S_{j}^{n-1}|Z_{j}^{n-1}S_{j}^{n-1}WS_{\mathcal{N}_{-j}}^{n}Z_{\mathcal{B}_{-j}}^{n})\\
 & -I(Y_{j}^{n-1}S_{j}^{n-1};Z_{jn}S_{jn}|Z_{j}^{n-1}S_{j}^{n-1}WS_{\mathcal{N}_{-j}}^{n}Z_{\mathcal{B}_{-j}}^{n})\\
 & +H(Y_{jn}|Y_{j}^{n-1}Z_{j}^{n}S_{j}^{n}WS_{\mathcal{N}_{-j}}^{n}Z_{\mathcal{B}_{-j}}^{n})\\
={} & H(Y_{j}^{n-1}S_{j}^{n-1}|Z_{j}^{n-1}S_{j}^{n-1}WS_{\mathcal{N}_{-j}}^{n}Z_{\mathcal{B}_{-j}}^{n})\\
 & -I(Y_{j}^{n-1}S_{j}^{n-1};Z_{jn}|Z_{j}^{n-1}S_{j}^{n-1}S_{jn}WS_{\mathcal{N}_{-j}}^{n}Z_{\mathcal{B}_{-j}}^{n})\\
 & +H(Y_{jn}|Y_{j}^{n-1}Z_{j}^{n}S_{j}^{n}WS_{\mathcal{N}_{-j}}^{n}Z_{\mathcal{B}_{-j}}^{n})\\
={} & H(Y_{j}^{n-1}S_{j}^{n-1}|Z_{j}^{n-1}S_{j}^{n-1}WS_{\mathcal{N}_{-j}}^{n}Z_{\mathcal{B}_{-j}}^{n})\\
 & -I(Y_{j}^{n-1}S_{j}^{n-1};Z_{jn}|Z_{j}^{n-1}S_{j}^{n-1}S_{jn}\in\{E_{j},E_{j}B_{j}\}WS_{\mathcal{N}_{-j}}^{n}Z_{\mathcal{B}_{-j}}^{n})\cdot\\
 & \quad\cdot\Pr\{S_{jn}\in\{E_{j},E_{j}B_{j}\}\}\\
 & +H(Y_{jn}|Y_{j}^{n-1}Z_{j}^{n}S^{n-1}W,S_{jn}=B_{j},S_{\mathcal{N}_{-j}}^{n}Z_{\mathcal{B}_{-j}}^{n})\Pr\{S_{jn}=B_{j}\}\\
 & +H(Y_{jn}|Y_{j}^{n-1}Z_{j}^{n}S^{n-1}W,S_{jn}=E_{j}B_{j},S_{\mathcal{N}_{-j}}^{n}Z_{\mathcal{B}_{-j}}^{n})\Pr\{S_{jn}=E_{j}B_{j}\}\\
={} & H(Y_{j}^{n-1}S_{j}^{n-1}|Z_{j}^{n-1}S_{j}^{n-1}WS_{\mathcal{N}_{-j}}^{n}Z_{\mathcal{B}_{-j}}^{n})\\
 & -I(Y_{j}^{n-1}S_{j}^{n-1};X_{jn}|Z_{j}^{n-1}S_{j}^{n-1}WS_{\mathcal{N}_{-j}}^{n}Z_{\mathcal{B}_{-j}}^{n})(1-\delta_{jE})\\
 & +H(X_{jn}|Y_{j}^{n-1}Z_{j}^{n-1}S_{j}^{n-1}WS_{\mathcal{N}_{-j}}^{n}Z_{\mathcal{B}_{-j}}^{n})(1-\delta_{j})\delta_{jE}\\
 & +H(X_{jn}|Y_{j}^{n-1}Z_{j}^{n-1}X_{jn}S_{j}^{n-1}WS_{\mathcal{N}_{-j}}^{n}Z_{\mathcal{B}_{-j}}^{n})(1-\delta_{j})(1-\delta_{jE})\\
={} & H(Y_{j}^{n-1}S_{j}^{n-1}|Z_{j}^{n-1}S_{j}^{n-1}WS_{\mathcal{N}_{-j}}^{n}Z_{\mathcal{B}_{-j}}^{n})\\
 & -I(Y_{j}^{n-1}S_{j}^{n-1};X_{jn}|Z_{j}^{n-1}S_{j}^{n-1}WS_{\mathcal{N}_{-j}}^{n}Z_{\mathcal{B}_{-j}}^{n})(1-\delta_{jE})\\
 & +H(X_{jn}|Y_{j}^{n-1}Z_{j}^{n-1}S_{j}^{n-1}WS_{\mathcal{N}_{-j}}^{n}Z_{\mathcal{B}_{-j}}^{n})(1-\delta_{j})\delta_{jE}
\end{align*}
All we needed was the independence property of $S_{jn}$. We can perform the same steps recursively
to obtain the result.
\end{IEEEproof}
The following lemma connects the available randomness with the random innovative (both to the
next node and Eve) packets that the transmitter can produce.
\begin{lem}
\label{lem:D}It is:

\[
H\left(U\right)\geq\left(1-\delta\delta_{E}\right)\sum_{i=1}^{n}H\left(X_{i}\mid Y^{i-1}Z^{i-1}S^{i-1}W\right)
\]
\end{lem}
\begin{IEEEproof}
\begin{eqnarray}
H\left(U\right) & \geq & H\left(U\mid W\right)\nonumber \\
 & \geq & I\left(U;Y^{n}Z^{n}S^{n}\mid W\right)\nonumber \\
 & = & \sum_{i=1}^{n}I\left(U;Y_{i}Z_{i}S_{i}\mid Y^{i-1}Z^{i-1}S^{i-1}W\right)\nonumber \\
 & = & \sum_{i=1}^{n}I\left(U;Y_{i}Z_{i}\mid Y^{i-1}Z^{i-1}S^{i-1}WS_{i}\right)\;\textrm{since \ensuremath{S_{i}}is ind. of \ensuremath{(UWY^{i-1}Z^{i-1}S^{i-1})}}\nonumber \\
 & = & \sum_{i=1}^{n}I\left(U;Y_{i}Z_{i}\mid Y^{i-1}Z^{i-1}S^{i-1}WS_{i}\in\left\{ B,E,EB\right\} \right)\cdot Pr\left\{ S_{i}\in\left\{ B,E,EB\right\} \right\} \nonumber \\
 & = & \left(1-\delta\delta_{E}\right)\sum_{i=1}^{n}I\left(U;X_{i}\mid Y^{i-1}Z^{i-1}S^{i-1}WS_{i}\in\left\{ B,E,EB\right\} \right)\nonumber \\
 & = & \left(1-\delta\delta_{E}\right)\sum_{i=1}^{n}I\left(U;X_{i}\mid Y^{i-1}Z^{i-1}S^{i-1}W\right)\;\textrm{since \ensuremath{S_{i}}is ind. of \ensuremath{(UWY^{i-1}Z^{i-1}S^{i-1}X_{i})}}\nonumber \\
 & = & \left(1-\delta\delta_{E}\right)\sum_{i=1}^{n}H\left(X_{i}\mid Y^{i-1}Z^{i-1}S^{i-1}W\right)-H\left(X_{i}\mid Y^{i-1}Z^{i-1}S^{i-1}WU\right)\nonumber \\
 & = & \left(1-\delta\delta_{E}\right)\sum_{i=1}^{n}H\left(X_{i}\mid Y^{i-1}Z^{i-1}S^{i-1}W\right)\;\textrm{due to (\ref{eq:def1_1})}.\label{eq:D}
\end{eqnarray}

\end{IEEEproof}
The next two theorems provide the converse for theorems \ref{thm:The-secret-key} and \ref{thm:The-secret-message}.
\begin{thm}
\label{thm:(Coverse of SK)}(Coverse for Secret Key)

\[
R_{SK}\leq\begin{cases}
\frac{D\delta_{E}(1-\delta)}{1-\delta\delta_{E}} & \; if\: D<1-\delta\delta_{E}\\
(1-\delta)\delta_{E} & \; if\: D\geq1-\delta\delta_{E}
\end{cases}.
\]
\end{thm}
\begin{IEEEproof}
It is,
\begin{eqnarray}
nR_{SK} & \leq & \log\left\lceil 2^{nR_{SK}}\right\rceil \nonumber \\
 & = & H(K)+o(n)\;\textrm{due to (\ref{eq:Uniformity})}\nonumber \\
 & = & H(K\mid Z^{n}S^{n})+I(K;Z^{n}S^{n})+o(n)\nonumber \\
 & \leq & H(K\mid Z^{n}S^{n})+o(n)\;\textrm{due to (\ref{eq:Security}})\nonumber \\
 & = & I(K;\hat{K}\mid Z^{n}S^{n})+H(K\mid\hat{K}Z^{n}S^{n})+o(n)\nonumber \\
 & \leq & I(K;\hat{K}\mid Z^{n}S^{n})+o(n)\;\textrm{due to (\ref{eq:ErrorZero}) and Fano ineq.}\nonumber \\
 & \leq & I(X^{n}S^{n};Y^{n}S^{n}\mid Z^{n}S^{n})+o(n)\nonumber \\
 & \leq & H(Y^{n}S^{n}\mid Z^{n}S^{n})+o(n)\label{eq: Rsk}
\end{eqnarray}
Where the second to last inequality is due to $K\rightarrow X^{n}S^{n}\rightarrow Y^{n}S^{n}\rightarrow\hat{K}$
being a Markov Chain (even when conditioned on $Z^{n}S^{n}$). And using lemma \ref{lem:C}
for $N=1$ and $\mathcal{B}=\{1\}$ we conclude (dropping the $j$ subscripts),
\begin{eqnarray*}
nR_{SK} & \leq & H(Y^{n}S^{n}\mid Z^{n}S^{n})+o(n)\\
 & \leq & (1-\delta)\delta_{E}\sum_{i=1}^{n}H\left(X_{i}\mid Y^{i-1}Z^{i-1}S^{i-1}\right)+o(n)\\
 & \leq & \frac{(1-\delta)\delta_{E}}{1-\delta\delta_{E}}H\left(U\right)+o(n)\;\textrm{due to Lemma \ref{lem:D}}.
\end{eqnarray*}

Thus:

\[
R_{SK}\leq\frac{(1-\delta)\delta_{E}}{1-\delta\delta_{E}}\underset{n\rightarrow\infty}{\liminf}\frac{H\left(U\right)}{n}
\]

The second inequality is direct application of the Maurer bound.\end{IEEEproof}
\begin{thm}
\label{thm:(Coverse-for-SM}(Coverse for Secret Message)
\end{thm}
\[
R_{SM}\leq\begin{cases}
\frac{\left(1-\delta\right)\delta_{E}}{1-\delta_{E}}D & \;\textrm{if \ensuremath{D\leq\frac{\left(1-\delta_{E}\right)\left(1-\delta\delta_{E}\right)}{1-\delta\delta_{E}^{2}}L\log q}}\\
\left(1-\delta\right)\delta_{E}\frac{1-\delta\delta_{E}}{1-\delta\delta_{E}^{2}}L\log q & \;\textrm{if \ensuremath{D>\frac{\left(1-\delta_{E}\right)\left(1-\delta\delta_{E}\right)}{1-\delta\delta_{E}^{2}}L\log q}}
\end{cases}.
\]

\begin{IEEEproof}
The the converse linear program is:

\begin{eqnarray*}
max & m\\
\frac{1-\delta_{E}}{1-\delta\delta_{E}}m & \leq & k\\
\frac{k}{(1-\delta)\delta_{E}}+\frac{m}{1-\delta} & \leq & L\log q\\
k & \leq & \frac{D\delta_{E}(1-\delta)}{1-\delta\delta_{E}}
\end{eqnarray*}

The first two equations where derived in \cite{czap2011secret} and the last is derived in
Theorem \ref{thm:(Coverse of SK)}. Solving this linear program, we come to the desired conclusion.
\end{IEEEproof}

\section*{Appendix B}

In this section we will prove Theorem \ref{thm:OuterBound}. We will construct a converse LP
which will be equivalent to the achievability LP and consequently will have the same optimal
value. After we derive the inequalities we make the following correspondance of terms:

\[
n\cdot d_{j}\leftrightarrow H(Y_{j}^{n}\mid S_{\mathcal{N}}^{n})
\]

\[
n\cdot k_{j}\leftrightarrow\delta_{jE}(1-\delta_{j})\sum_{i=1}^{n}H(X_{ji}|Y_{j}^{i-1}Z_{j}^{i-1}S_{j}^{i-1}WS_{\mathcal{N}_{-j}}^{n})
\]

\[
D_{j}\leftrightarrow\underset{n\rightarrow\infty}{\liminf}\frac{H\left(U_{j}\right)}{n}
\]

This means that we forget the meaning of these information theoretic measures and we only use
the fact that they are non-negative variables. Some terms correspond to more than one variables.
This can only increase the value of the LP. 

The next three lemmas are generalizations of the equivalent results in \cite{czap2011secret}.
\begin{lem}
\textup{\label{R-1}It is $\forall j\in\mathcal{N}$,}

\[
n\cdot m\leq(1-\delta_{j})\sum_{i=1}^{n}I(W;X_{ji}|Y_{j}^{i-1}S_{j}^{i-1}S_{\mathcal{N}_{-j}}^{n}).
\]
\end{lem}
\begin{IEEEproof}
Let,
\begin{eqnarray*}
n\cdot m & \leq & I(W;Y_{N}^{n}S_{\mathcal{N}}^{n})\\
 & \leq & I(W;Y_{j}^{n}S_{\mathcal{N}}^{n})\;\textrm{since it is a Markov chain }\\
 & = & I(W;Y_{j}^{n}S_{j}^{n}|S_{\mathcal{N}_{-j}}^{n})\;\textrm{since \ensuremath{S_{\mathcal{N}_{-j}}^{n}}is indep. of \ensuremath{W}}\\
 & = & \sum_{i=1}^{n}I(W;Y_{ji}S_{ji}|Y_{j}^{i-1}S_{j}^{i-1}S_{\mathcal{N}_{-j}}^{n})\\
 & = & \sum_{i=1}^{n}I(W;Y_{ji}|Y_{j}^{i-1}S_{j}^{i-1}S_{\mathcal{N}_{-j}}^{n}S_{ji})\;\textrm{since \ensuremath{S_{ji}}is indep. of \ensuremath{(WY_{j}^{i-1}S_{j}^{i-1}S_{\mathcal{N}_{-j}}^{n})}}\\
 & = & \sum_{i=1}^{n}I(W;X_{ji}|Y_{j}^{i-1}S_{j}^{i-1}S_{\mathcal{N}_{-j}}^{n}S_{ji}\in\{B_{j},E_{j}B_{j}\})\cdot\Pr\{S_{ji}\in\{B_{j},E_{j}B_{j}\}\}\\
 & = & (1-\delta_{j})\sum_{i=1}^{n}I(W;X_{ji}|Y_{j}^{i-1}S_{j}^{i-1}S_{\mathcal{N}_{-j}}^{n}).
\end{eqnarray*}
\end{IEEEproof}
\begin{lem}
\textup{\label{e-2}It is $\forall\mathcal{N_{\textrm{E}}}\underset{V}{\subset}\mathcal{N}$
and $\forall j\in\mathcal{N}_{E}$,}

\begin{eqnarray*}
\sum_{i=1}^{n}I(W;X_{ji}|Z_{j}^{i-1}S_{j}^{i-1}S_{\mathcal{N}_{-j}}^{n}Z_{[j-1]_{E}}^{n}) & < & \frac{\epsilon}{1-\delta_{jE}}
\end{eqnarray*}
\end{lem}
\begin{IEEEproof}
Let,

\begin{eqnarray*}
\epsilon & > & I(W;Z_{[j]_{E}}^{n}S_{\mathcal{N}}^{n})\\
 & \geq & I(W;Z_{j}^{n}S_{\mathcal{N}}^{n}|Z_{[j-1]_{E}}^{n})\\
 & = & I(W;Z_{j}^{n}S_{j}^{n}|S_{\mathcal{N}_{-j}}^{n}Z_{[j-1]_{E}}^{n})\;\textrm{since \ensuremath{S_{\mathcal{N}_{-j}}^{n}} is indep. of \ensuremath{W}}\textrm{and \ensuremath{Z_{[j-1]_{E}}^{n}}}\\
 & = & \sum_{i=1}^{n}I(W;Z_{ji}S_{ji}|Z_{j}^{i-1}S_{j}^{i-1}S_{\mathcal{N}_{-j}}^{n}Z_{[j-1]_{E}}^{n})\\
 & = & \sum_{i=1}^{n}I(W;Z_{ji}|Z_{j}^{i-1}S_{j}^{i-1}S_{\mathcal{N}_{-j}}^{n}S_{ji}Z_{[j-1]_{E}}^{n})\;\textrm{since \ensuremath{S_{ji}}is indep. of \ensuremath{(WY_{j}^{i-1}S_{j}^{i-1}S_{\mathcal{N}_{-j}}^{n}Z_{[j-1]_{E}}^{n})}}\\
 & = & \sum_{i=1}^{n}I(W;X_{ji}|Z_{j}^{i-1}S_{j}^{i-1}S_{\mathcal{N}_{-j}}^{n}Z_{[j-1]_{E}}^{n}S_{ji}\in\{E_{j},E_{j}B_{j}\})\cdot\Pr\{S_{i}\in\{E_{j},E_{j}B_{j}\}\}\\
 & = & (1-\delta_{jE})\sum_{i=1}^{n}I(W;X_{ji}|Z_{j}^{i-1}S_{j}^{i-1}S_{\mathcal{N}_{-j}}^{n}Z_{[j-1]_{E}}^{n})
\end{eqnarray*}
\end{IEEEproof}
\begin{lem}
\textup{\label{dd-2}It is $\forall\mathcal{N_{\textrm{E}}}\underset{V}{\subset}\mathcal{N}$
and $\forall j\in\mathcal{N}_{E}$,}

\begin{eqnarray*}
\sum_{i=1}^{n}I(W;X_{ji}|Y_{j}^{i-1}Z_{j}^{i-1}S_{j}^{i-1}S_{\mathcal{N}_{-j}}^{n}Z_{[j-1]_{E}}^{n}) & \geq & \frac{n\cdot m}{1-\delta_{j}\delta_{jE}}
\end{eqnarray*}
\end{lem}
\begin{IEEEproof}
Let,

\begin{eqnarray*}
n\cdot m & \leq & I(W;Y_{N}^{n}S_{\mathcal{N}}^{n})\\
 & \leq & I(W;Y_{N}^{n}Z_{[j]_{E}}^{n}S_{\mathcal{N}}^{n})\\
 & \leq & I(W;Y_{j}^{n}Z_{[j]_{E}}^{n}S_{\mathcal{N}}^{n})\;\textrm{since it is a Markov chain }\\
 & = & I(W;Y_{j}^{n}Z_{[j]_{E}}^{n}S_{j}^{n}|S_{\mathcal{N}_{-j}}^{n})\;\textrm{since \ensuremath{S_{\mathcal{N}_{-j}}^{n}}is indep. of \ensuremath{W}}\\
 & = & I(W;Y_{j}^{n}Z_{j}^{n}S_{\mathcal{N}}^{n}|S_{\mathcal{N}_{-j}}^{n}Z_{[j-1]_{E}}^{n})+\epsilon\;\textrm{due to (\ref{eq:def1_5})}\\
 & = & \sum_{i=1}^{n}I(W;Y_{ji}Z_{ji}S_{ji}|Y_{j}^{i-1}Z_{j}^{i-1}S_{j}^{i-1}S_{\mathcal{N}_{-j}}^{n}Z_{[j-1]_{E}}^{n})+\epsilon\\
 & = & \sum_{i=1}^{n}I(W;Y_{ji}Z_{ji}|Y_{j}^{i-1}Z_{j}^{i-1}S_{j}^{i-1}S_{\mathcal{N}_{-j}}^{n}Z_{[j-1]_{E}}^{n}S_{ji})+\epsilon\;\textrm{since \ensuremath{S_{ji}}is indep. of \ensuremath{(WY_{j}^{i-1}Z_{j}^{i-1}S_{j}^{i-1}S_{\mathcal{N}_{-j}}^{n}Z_{[j-1]}^{n})}}\\
 & = & \sum_{i=1}^{n}I(W;X_{ji}|Y_{j}^{i-1}Z_{j}^{i-1}S_{j}^{i-1}S_{\mathcal{N}_{-j}}^{n}Z_{[j-1]_{E}}^{n}S_{ji}\in\{B_{j},E_{j}B_{j},E_{j}\})\cdot\Pr\{S_{ji}\in\{B_{j},E_{j}B_{j},E_{j}\}\}+\epsilon\\
 & = & (1-\delta_{j}\delta_{jE})\sum_{i=1}^{n}I(W;X_{ji}|Y_{j}^{i-1}Z_{j}^{i-1}S_{j}^{i-1}S_{\mathcal{N}_{-j}}^{n}Z_{[j-1]_{E}}^{n})+\epsilon.
\end{eqnarray*}

\end{IEEEproof}
\textbf{It is $\forall\mathcal{N_{\textrm{E}}}\underset{V}{\subset}\mathcal{N}$:}

\textbf{For the first constraint} $\forall j\in\mathcal{N_{\textrm{}}}$\textbf{:}

\begin{eqnarray*}
n\cdot k_{j}^{\mathcal{N_{\textrm{E}}}}-n\cdot d_{j}^{\mathcal{N_{\textrm{E}}}} & = & (1-\delta_{j})\delta_{jE}\sum_{i=1}^{n}H(X_{ji}|Y_{j}^{i-1}Z_{j}^{i-1}S_{j}^{i-1}WS_{\mathcal{N}_{-j}}^{n}Z_{[j-1]_{E}}^{n})-H\left(Y_{j}^{n}\mid WS_{\mathcal{N}}^{n}Z_{[j]_{E}}^{n}\right)\\
 & \geq & (1-\delta_{jE})\sum_{i=1}^{n}I(Y_{j}^{i-1}S_{j}^{i-1};X_{ji}|Z_{j}^{i-1}S_{j}^{i-1}WS_{\mathcal{N}_{-j}}^{n}Z_{[j-1]_{E}}^{n})\:\textrm{due to Lemma }\ref{lem:C}\\
 & = & (1-\delta_{jE})\sum_{i=1}^{n}H(X_{i}|Z^{i-1}S^{i-1}WS_{\mathcal{N}_{-j}}^{n}Z_{[j-1]_{E}}^{n})-H(X_{i}|Y^{i-1}Z^{i-1}S^{i-1}WS_{\mathcal{N}_{-j}}^{n}Z_{[j-1]_{E}}^{n})\\
 & = & (1-\delta_{jE})\sum_{i=1}^{n}H(X_{i}|Z^{i-1}S^{i-1}S_{\mathcal{N}_{-j}}^{n}Z_{[j-1]_{E}}^{n})-H(X_{i}|Y^{i-1}Z^{i-1}S^{i-1}WS_{\mathcal{N}_{-j}}^{n}Z_{[j-1]_{E}}^{n})\\
 &  & -I(W;X_{i}|Z^{i-1}S^{i-1}S_{\mathcal{N}_{-j}}^{n}Z_{[j-1]_{E}}^{n})\\
 & \geq & (1-\delta_{jE})\sum_{i=1}^{n}H(X_{i}|Y^{i-1}Z^{i-1}S^{i-1}S_{\mathcal{N}_{-j}}^{n}Z_{[j-1]_{E}}^{n})-H(X_{i}|Y^{i-1}Z^{i-1}S^{i-1}WS_{\mathcal{N}_{-j}}^{n}Z_{[j-1]_{E}}^{n})\\
 &  & -I(X_{i};W|Z^{i-1}S^{i-1}S_{\mathcal{N}_{-j}}^{n}Z_{[j-1]_{E}}^{n})\\
 & = & (1-\delta_{jE})\sum_{i=1}^{n}I(W;X_{ji}|Y_{j}^{i-1}Z_{j}^{i-1}S_{j}^{i-1}S_{\mathcal{N}_{-j}}^{n}Z_{[j-1]_{E}}^{n})-I(W;X_{ji}|Z_{j}^{i-1}S_{j}^{i-1}S_{\mathcal{N}_{-j}}^{n}Z_{[j-1]_{E}}^{n})\\
 & \geq & \frac{1-\delta_{jE}}{1-\delta_{j}\delta_{jE}}n\cdot m-\epsilon\;\textrm{by Lemmas \ref{e-2} and \ref{dd-2}}
\end{eqnarray*}

\textbf{For the second constraint} $\forall j\in\mathcal{N_{\textrm{}}}$\textbf{:}

\begin{eqnarray*}
n\cdot m & \leq & (1-\delta_{j})\sum_{i=1}^{n}I(W;X_{ji}|Y_{j}^{i-1}S_{j}^{i-1}S_{\mathcal{N}_{-j}}^{n})\;\textrm{by Lemma \ref{R-1} }\\
 & = & (1-\delta_{j})\sum_{i=1}^{n}H(X_{ji}|Y_{j}^{i-1}S_{j}^{i-1}S_{\mathcal{N}_{-j}}^{n})-H(X_{ji}|Y_{j}^{i-1}S_{j}^{i-1}S_{\mathcal{N}_{-j}}^{n}W)\\
 & \leq & (1-\delta_{j})nL\log q-(1-\delta_{j})\sum_{i=1}^{n}H(X_{ji}|Y_{j}^{i-1}S_{j}^{i-1}S_{\mathcal{N}_{-j}}^{n}W)\\
 & \leq & (1-\delta_{j})nL\log q-(1-\delta_{j})\sum_{i=1}^{n}H(X_{ji}|Y_{j}^{i-1}S_{j}^{i-1}Z_{j}^{i-1}S_{\mathcal{N}_{-j}}^{n}Z_{[j-1]_{E}}^{n}W)\\
 & = & (1-\delta_{j})nL\log q-\frac{n\cdot k_{j}^{\mathcal{N_{\textrm{E}}}}}{\delta_{jE}}
\end{eqnarray*}

\textbf{For the third constraint} $\forall j\in\mathcal{N_{\textrm{}}}$\textbf{:} (let $Y_{0}^{n}=c$,
a constant)

\begin{eqnarray*}
 &  & H\left(Y_{j-1}^{n}\mid WS_{\mathcal{N}}^{n}Z_{[j-1]_{E}}^{n}\right)+H(U_{j-1})\\
 & = & H\left(Y_{j-1}^{n}U_{j-1}\mid WS_{\mathcal{N}}^{n}Z_{[j-1]_{E}}^{n}\right)\;\textrm{since \ensuremath{U_{j-1}}is ind. of \ensuremath{(Y_{j-1}^{n}WS_{\mathcal{N}}^{n}Z_{[j-1]_{E}}^{n})}}\\
 & = & H\left(Y_{j-1}^{n}U_{j-1}\mid WS_{\mathcal{N}_{-j}}^{n}Z_{[j-1]_{E}}^{n}\right)\;\textrm{since \ensuremath{S_{j}^{n}}is ind. of \ensuremath{(Y_{j-1}^{n}WU_{j-1}S_{\mathcal{N}_{-j}}^{n}Z_{[j-1]_{E}}^{n})}}\\
 & \geq & I\left(Y_{j-1}^{n}U_{j-1};Y_{j}^{n}Z_{j}^{n}S_{j}^{n}\mid WS_{\mathcal{N}_{-j}}^{n}Z_{[j-1]_{E}}^{n}\right)\\
 & = & \sum_{i=1}^{n}I\left(Y_{j-1}^{n}U_{j-1};Y_{ji}Z_{ji}S_{ji}\mid Y_{j}^{i-1}Z_{j}^{i-1}S_{j}^{i-1}WS_{\mathcal{N}_{-j}}^{n}Z_{[j-1]_{E}}^{n}\right)\\
 & = & \sum_{i=1}^{n}I\left(Y_{j-1}^{n}U_{j-1};Y_{ji}Z_{ji}\mid Y_{j}^{i-1}Z_{j}^{i-1}S_{j}^{i-1}WS_{\mathcal{N}_{-j}}^{n}Z_{[j-1]_{E}}^{n}S_{ji}\right)\;\textrm{since \ensuremath{S_{ji}}is ind. of \ensuremath{(Y_{j-1}^{n}U_{j-1}WY^{i-1}Z^{i-1}S^{i-1}S_{\mathcal{N}_{-j}}^{n}Z_{[j-1]_{E}}^{n})}}\\
 & = & \sum_{i=1}^{n}I\left(Y_{j-1}^{n}U_{j-1};Y_{ji}Z_{ji}\mid Y_{j}^{i-1}Z_{j}^{i-1}S_{j}^{i-1}WS_{\mathcal{N}_{-j}}^{n}Z_{[j-1]_{E}}^{n}S_{ji}\in\left\{ B_{j},E_{j},E_{j}B_{j}\right\} \right)\cdot Pr\left\{ S_{i}\in\left\{ B_{j},E_{j},E_{j}B_{j}\right\} \right\} \\
 & = & \left(1-\delta_{j}\delta_{jE}\right)\sum_{i=1}^{n}I\left(Y_{j-1}^{n}U_{j-1};X_{ji}\mid Y_{j}^{i-1}Z_{j}^{i-1}S_{j}^{i-1}WS_{\mathcal{N}_{-j}}^{n}Z_{[j-1]_{E}}^{n}S_{ji}\in\left\{ B_{j},E_{j},E_{j}B_{j}\right\} \right)\\
 & = & \left(1-\delta_{j}\delta_{jE}\right)\sum_{i=1}^{n}I\left(Y_{j-1}^{n}U_{j-1};X_{ji}\mid Y_{j}^{i-1}Z_{j}^{i-1}S_{j}^{i-1}Z_{[j-1]_{E}}^{n}WS_{\mathcal{N}_{-j}}^{n}\right)\\
 &  & \textrm{since \ensuremath{S_{ji}}is ind. of \ensuremath{(Y_{j-1}^{n}U_{j-1}X_{ji}WY_{j}^{i-1}Z_{j}^{i-1}S_{j}^{i-1}S_{\mathcal{N}_{-j}}^{n}Z_{[j-1]_{E}}^{n})}}\\
 & = & \left(1-\delta_{j}\delta_{jE}\right)\sum_{i=1}^{n}H\left(X_{ji}\mid Y_{j}^{i-1}Z_{j}^{i-1}S_{j}^{i-1}WS_{\mathcal{N}_{-j}}^{n}Z_{[j-1]_{E}}^{n}\right)-H\left(X_{i}\mid Y_{j}^{i-1}Z_{j}^{i-1}S_{j}^{i-1}WY_{j-1}^{n}U_{j-1}S_{\mathcal{N}_{-j}}^{n}Z_{[j-1]_{E}}^{n}\right)\\
 & = & \left(1-\delta_{j}\delta_{jE}\right)\sum_{i=1}^{n}H\left(X_{ji}\mid Y_{j}^{i-1}Z_{j}^{i-1}S_{j}^{i-1}WS_{\mathcal{N}_{-j}}^{n}Z_{[j-1]_{E}}^{n}\right)\;\textrm{due to (\ref{eq:def1_1})}\\
 & = & \frac{1-\delta_{j}\delta_{jE}}{(1-\delta_{j})\delta_{jE}}n\cdot k_{j}^{\mathcal{N_{\textrm{E}}}}
\end{eqnarray*}

Thus:

\begin{eqnarray*}
k_{j}^{\mathcal{N_{\textrm{E}}}} & \leq & \frac{(1-\delta_{j})\delta_{jE}}{1-\delta_{j}\delta_{jE}}\left(d_{j-1}+\underset{n\rightarrow\infty}{\liminf}\frac{H\left(U_{j-1}\right)}{n}\right)\\
 & = & \frac{(1-\delta_{j})\delta_{jE}}{1-\delta_{j}\delta_{jE}}\left(d_{j-1}+D_{j-1}\right)
\end{eqnarray*}

\textbf{For the fourth constraint} $\forall j\in\mathcal{N_{\textrm{}}}-\mathcal{N_{\textrm{E}}}$\textbf{:}
\begin{eqnarray*}
n\cdot m+n\cdot d_{j}^{\mathcal{N_{\textrm{E}}}} & \leq & I(W;Y_{j}^{n}S_{\mathcal{N}}^{n})+H(Y_{j}^{n}\mid WS_{\mathcal{N}}^{n}Z_{[j]_{E}}^{n})\\
 & = & I(W;Y_{j}^{n}\mid S_{\mathcal{N}}^{n})+H(Y_{j}^{n}\mid WS_{\mathcal{N}}^{n}Z_{[j]_{E}}^{n})\;\textrm{since \ensuremath{S_{\mathcal{N}}^{n}}is indep. of \ensuremath{W}}\\
 & \leq & I(W;Y_{j}^{n}\mid S_{\mathcal{N}}^{n}Z_{[j]_{E}}^{n})+H(Y_{j}^{n}\mid WS_{\mathcal{N}}^{n}Z_{[j]_{E}}^{n})+\epsilon\;\textrm{due to (\ref{eq:def1_5})}\\
 & = & H(Y_{j}^{n}\mid S_{\mathcal{N}}^{n}Z_{[j]_{E}}^{n})+\epsilon\\
 & \leq & (1-\delta_{j})nL\log q+\epsilon
\end{eqnarray*}

\textbf{For the fifth constraint} $\forall j\in\mathcal{N_{\textrm{}}}-\mathcal{N_{\textrm{E}}}$\textbf{:}

\begin{eqnarray*}
n\cdot d_{j}^{\mathcal{N_{\textrm{E}}}} & = & H(Y_{j}^{n}\mid WS_{\mathcal{N}}^{n}Z_{[j]_{E}}^{n})\\
 & = & H(Y_{j}^{n}\mid WS_{\mathcal{N}}^{n}Z_{[j-1]_{E}}^{n})\textrm{ since }j\in\mathcal{N_{\textrm{}}}-\mathcal{N_{\textrm{E}}}\\
 & = & H(Y_{j}^{n}Z_{j}^{n}\mid WS_{\mathcal{N}}^{n}Z_{[j-1]_{E}}^{n})\\
 & \leq & H(Y_{j-1}^{n}U_{j-1}\mid WS_{\mathcal{N}}^{n}Z_{[j-1]_{E}}^{n})\;\textrm{due to Markovity}\\
 & = & H\left(Y_{j-1}^{n}\mid WS_{\mathcal{N}}^{n}Z_{[j-1]_{E}}^{n}\right)+H(U_{j-1})\\
 & = & n\cdot d_{j-1}^{\mathcal{N_{\textrm{E}}}}+n\cdot D_{j-1}
\end{eqnarray*}

Thus:

\begin{eqnarray*}
d_{j}^{\mathcal{N_{\textrm{E}}}} & \leq & d_{j-1}^{\mathcal{N_{\textrm{E}}}}+\underset{n\rightarrow\infty}{\liminf}\frac{H(U_{j-1})}{n}\\
 & = & d_{j-1}^{\mathcal{N_{\textrm{E}}}}+D_{j-1}
\end{eqnarray*}

\end{document}